\documentclass{article}
\usepackage{graphicx}
\usepackage{cite} 
\usepackage{epsfig}

\begin{document}

\title{
Multi-scale analysis of lung computed tomography images
} 

\author{Ilaria Gori$^{ab*}$,
Francesco Bagagli$^c$, Maria Evelina Fantacci$^{ac}$,\\ 
Alessandro Preite Martinez$^d$, Alessandra Retico$^a$, Ivan De Mitri$^{e}$,\\ 
Stefania Donadio$^{f}$, Christian Fulcheri$^{g}$, Gianfranco Gargano$^{h}$, \\
Rosario Magro$^i$, Matteo Santoro$^j$, Simone Stumbo$^{kl}$\\
\\
\llap{$^a$}{\small Istituto Nazionale di Fisica Nucleare (INFN), Sezione di Pisa,} \\
{\small Largo Pontecorvo 3, 56127 Pisa, Italy}\\
\llap{$^b$}{\small Bracco Imaging S.p.A. Via E. Folli 50, 20134 Milano, Italy}\\
\llap{$^c$}{\small Dipartimento di Fisica, Universit\`a di Pisa, Largo Pontecorvo 3,}\\
{\small 56127 Pisa, Italy}\\
\llap{$^d$}{\small Centro Studi e Ricerche Enrico Fermi, Via Panisperna 89/A,}\\ 
{\small 00184 Roma, Italy}\\
\llap{$^e$}{\small Dipartimento di Fisica, Universit\`a del Salento, and INFN,}\\ 
{\small Sezione di Lecce, Via per Arnesano, 73100, Lecce, Italy}\\
\llap{$^f$}{\small Dipartimento di Fisica, Universit\`a di Genova, and INFN,} \\
{\small Sezione di Genova, Via Dodecaneso 33, 16146, Genova, Italy}\\
\llap{$^g$}{\small Dipartimento di Fisica Sperimentale, Universit\`a di Torino and INFN,}\\ 
{\small  Sezione di Torino, Via P. Giuria 1, 10125, Torino, Italy}\\
\llap{$^h$}{\small Dipartimento di Fisica, Universit\`a di Bari, and INFN,}\\ 
{\small Sezione di Bari, Via Amendola 173, 70126, Bari, Italy}\\
\llap{$^i$}{\small Dipartimento di Fisica e Tecnologie Relative, Universit\`a di Palermo, }\\
{\small Viale delle Scienze, Edificio 18, 90128, Palermo, Italy} \\
\llap{$^j$}{\small Dipartimento di Scienze Fisiche, Universit\`a Federico II di Napoli, }\\
{\small via Cintia-Complesso Monte S.Angelo, 80126, Napoli, Italy }\\
\llap{$^k$}{\small Struttura Dipartimentale di Matematica e Fisica, Universit\`a di Sassari, }\\
{\small Via Vienna 2, 07100, Sassari, Italy}\\
\llap{$^l$}{\small INFN, Sezione di Cagliari, Cittadella Universitaria di Monserrato, }\\
{\small Casella Postale 170, 09042, Monserrato (CA), Italy}\\
\llap{$^*$}{\small E-mail: ilaria.gori@pi.infn.it}
}

\date{}
\maketitle

\abstract{
A computer-aided detection (CAD) system for the identification of lung internal no\-dules in low-dose multi-detector helical Computed Tomography (CT) images was developed in the framework of the MAGIC-5 project. The three modules of our lung CAD system, a segmentation algorithm for lung internal region identification, a multi-scale dot-enhancement filter for nodule candidate selection and a multi-scale neural technique for false positive finding reduction, are described. The results obtained on a dataset of low-dose and thin-slice CT scans are shown in terms of free response receiver operating characteristic (FROC) curves and discussed.

{\bf \small keywords: Computed Tomography (CT), Computer-Aided Detection (CAD)}


\section{INTRODUCTION}

Lung cancer is the leading cause of cancer-related mortality in developed countries~\cite{cancer,cancer_stat_US}. Only 10--15\% of all men and women diagnosed with lung cancer live five years after diagnosis~\cite{cancer_stat_US,cancer_stat_EU} and no significant improvement has occurred in the last 20 years~\cite{Singh}. Early-stage cancer is a\-symptomatic, so more than 70\% of patients diagnosed with lung cancer are in the advanced stages of the disease, when it's too late for effective treatments~\cite{Ihde}. 
However the five-year survival rate for people who are diagnosed with early-stage lung cancer (stage I) can reach 70\%~\cite{Nesbitt}.\\
In this scenario, the implementation of screening programs for the asymptomatic high-risk population is an approach that is being tried to reduce the mortality rate of lung cancer. It was proved that screening programs with X-ray radiography don't lead to a reduction of the mortality rate~\cite{Frost,Melamed,Fontana,Marcus}, due to the low sensitivity of this technique in the identification of small, early-stage cancers.\\
Lung cancer most commonly manifests itself with the formation of non-calcified pulmonary no\-dules. Computed Tomography (CT) is proved to be the best imaging modality for the detection of small pulmonary nodules, particularly since the introduction of the multi-detector-row and helical CT technologies~\cite{Diederich,Kaneko,Sone}. \\
Therefore CT-based screening programs are regarded as a promising technique for detecting small, early-stage lung cancers~\cite{Itoh,Henschke}. In CT-based screening protocols, low-dose settings are required, since the examined population is asymptomatic, and therefore potentially healthy. However, low-dose images are noisier than standard-dose ones, so it's more difficult to identify small nodules when low-dose settings are used. Moreover, the amount of data that need to be interpreted in CT examinations can be very large, especially in screening programs, when a thin slice thickness is usually used, thus generating up to about 300 two-dimensional images per scan. Computer-Aided Detection (CAD) could support radiologists in the identification of small, early-stage pathological objects in screening CT scans. \\
The MAGIC-5 project~\cite{magic5} aims at developing CAD software systems for Medical Applications on distributed databases by means of a GRID Infrastructure Connection approach~\cite{GRID}. In particular, MAGIC-5 researchers work on mammographic images for breast cancer detection, NMR--SPECT--PET images for the diagnosis of the Alzheimer disease and Computed Tomography images for lung cancer identification. The CAD system we propose for small pulmonary internal nodule identification was developed and validated on a set of images acquired from the Pisa centre of the First Italian Randomized Controlled Trial (ITALUNG-CT), recently started in order to study the potential impact of screening on a high-risk population using low-dose helical CT~\cite{screening,italung2}. \\
The CAD system is a three steps procedure: a segmentation algorithm identifies the lung internal region, then a multi-scale dot-enhancement filter provides a list of nodule candidates and finally a multi-scale neural network-based classification module reduces the number of false positive findings per scan.\\

\section{THE LUNG CT DATABASE}
\label{subsubsect_DB}
A low-dose lung CT database was acquired from the Pisa centre of the ITALUNG-CT trial, the First Italian Randomized Controlled Trial for the screening of lung cancer~\cite{screening,italung2}. The CT scans are acquired with a 4-slice spiral CT scanner according to a low-dose protocol (screening setting: 140~kV, 20 mA, mean equivalent dose 0.6 mSv), with 1.25-mm slice collimation.\\
Each scan is a sequence of slices stored in DICOM (Digital Imaging and COmmunications in Medicine) format. The reconstructed slice thickness is 1 mm. The average number of slices per scan is about 300 with a 512$\times$512 pixel matrix, a pixel size ranging from 0.53 to 0.74 mm and 12~bit grey levels in Hounsfield units (HU).\\
The pathological structures to be automatically detected by a CAD system are non calcified nodules. Such nodules can be divided into three main categories, depending on their location in the lung: internal nodules, fully contained in the lung parenchyma, sub-pleural nodules, originated inside the lung parenchyma but adjacent or connected to the pleura, and pleural nodules, originated from the pleura and grown toward the lung parenchyma.\\
The identification of internal, sub-pleural and pleural nodules requires dedicated procedures, due to their different location and shape (see figure~\ref{fig:noduli3tipi}). Our preliminary results on sub-pleural and pleural nodule identification can be found in~\cite{SPIE,CARSparnian}.\\
In this study, only internal nodule identification is considered.
\begin{figure}[ht]
\begin{center}
\includegraphics[width=.9\textwidth]{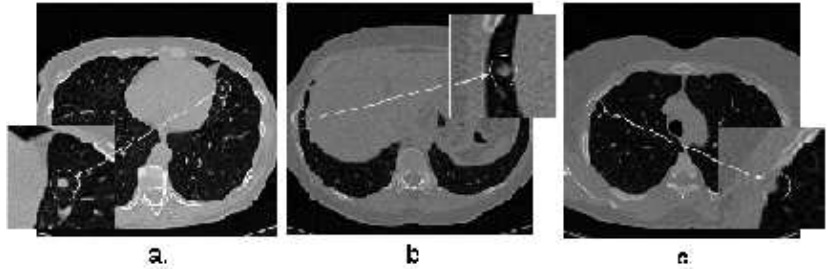}
\end{center}
\caption{Examples of small pulmonary nodules: a) internal nodule; b) sub-pleural nodule; c) pleural nodule.}
\label{fig:noduli3tipi}
\end{figure}
\noindent According to ITALUNG-CT screening protocol, in the baseline CT examinations radiologists mark the nodules with a diameter greater than 5 mm, which can be a sign of the presence of lung cancer at an early stage, thus allowing an early diagnosis of this disease. Once identified, nodules are kept under control by means of follow up CT examinations, in which the nodules with diameters between 3 and 5 mm should also be marked, but only if newly formed. Only nodules with diameters greater than 8-10 mm are subjected to further examinations like Positron Emission Tomography (PET) or biopsy.\\
The CT database acquired for this study is constituted by baseline examinations, so our goal is to develop a CAD system for the identification of internal nodules with diameters greater than 5 mm. Two experienced radiologists have selected the structures of interest from the so far collected database of 39 CT scans. This task has been carried out by means of a dedicated visualization and annotation tool developed in the framework of the MAGIC-5 collaboration. The resulting dataset consists of 75 solid internal nodules with diameters in the 5--12~mm range. The maximum value is 12 mm, but 96\% of the nodules have diameters ranging from 5 to 8 mm. Examples of solid internal nodules extracted from our dataset are shown in figure~\ref{fig:nodule_examples}: they may have CT values in the same range of those of blood vessels and airway walls and may be strongly connected to them.\\
Some internal ground-glass opacities had also been selected by radiologists, but their number was too small to allow a dedicated analysis; therefore this type of pathological objects was excluded from our target list.

\begin{figure}[ht]
\begin{center}
\includegraphics[width=.9\textwidth]{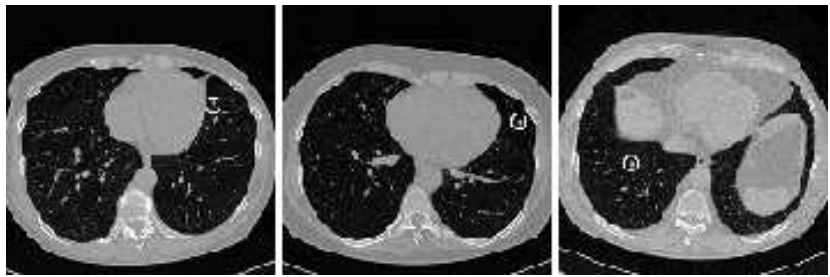}
\end{center}
\caption{Examples of internal small pulmonary nodules.}
\label{fig:nodule_examples}
\end{figure}

\section{THE CAD SYSTEM}

As explained in the Introduction, our CAD system consists in three main modules: first of all, the lung internal region is identified by means of a purposely built segmentation algorithm. As a second step, nodule candidates are detected using a multi-scale 3D  filter enhancing spherically-shaped objects. Finally, a multi-scale voxel-based neural technique is implemented to reduce the amount of false positive findings per scan. 

\subsection{Lung internal region segmentation}

A lung volume 3D segmentation algorithm was implemented according to the procedure proposed in~\cite{ITK}. First of all, to separate the low-intensity lung parenchyma from the high-intensity surrounding tissue (fat tissue and bones), the voxel intensities are thresholded at a fixed value (-400~HU) \cite{ITK,kemerink}. Then, in order to discard all the regions not belonging to the lungs, the connected lung regions for the left and right lungs are selected starting with a seed point inside each lung. At this stage vessels and airway walls are not included in the segmented lung. Finally, a combination of the erosion and dilation morphological operations, known as \emph{rolling-ball algorithm} \cite{rolling}, is applied. The rolling-ball operator uses a spherical kernel, having the effect of including in the segmented lung all the vessels and all the airway walls smaller than the ball size: a radius of 10 voxels for the spherical kernel was chosen in order to include all the objects within our nodule dimension target. An example of the various stages of the lung volume segmentation algorithm is shown in figure~\ref{fig:segm}.

\begin{figure}[ht]
\begin{center}
\includegraphics[width=\textwidth]{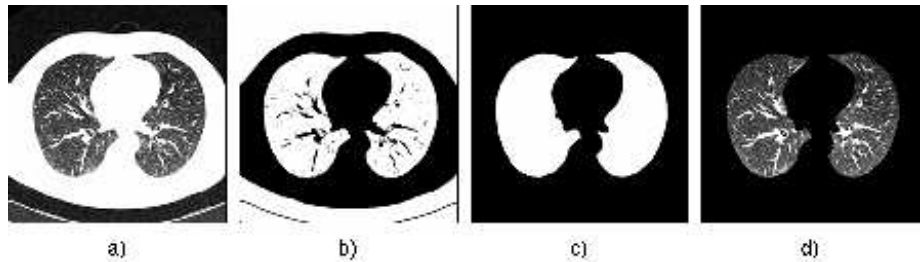}
\end{center}
\caption{Lung volume 3D segmentation algorithm (a slice of the full 3D volume is shown): a) algorithm input (the original CT scan); b) scan after thresholding; c) scan after selection of connected regions and rolling ball algorithm; d) algorithm output (segmented lung volume).}
\label{fig:segm}
\end{figure}
\noindent Since our goal is internal nodule research only, the lung volume identified by the segmentation algorithm is eroded on the external side by a 2.5 mm-thick band, so as to define the \emph{lung internal region}, where internal nodules with a diameter greater than 5 mm are expected to be found. In figure~\ref{fig:segm_banda} an example of lung internal region is shown, compared with the lung volume identified by simply implementing the segmentation algorithm described above.

\begin{figure}[ht]
\begin{center}
\includegraphics[width=.6\textwidth]{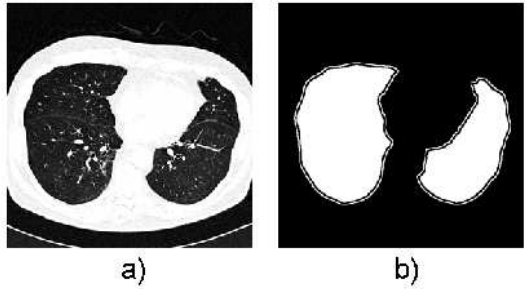}
\end{center}
\caption{a) Original volume; b) lung volume obtained by the segmentation algorithm (outer contours) and lung internal region.}
\label{fig:segm_banda}
\end{figure}

\subsection{Nodule candidate identification}

The automated nodule candidate detection should be characterized by a sensitivity value close to 100\%, in order to avoid setting an \emph{a priori} upper bound to the CAD system performance. To this aim, we followed the approach proposed in~\cite{Li}. Lung nodules are modeled as spherical objects with a Gaussian profile and the 3D matrix of data is filtered with a \emph{filter function} $z_{dot}$ 
built to discriminate between spherical objects and objects with planar or elongated shapes. In particular, $z_{dot}$ is defined as follows from the eigenvalues of the Hessian matrix of each voxel:
\begin{eqnarray*}
z_{dot}(\lambda_1,\lambda_2,\lambda_3)= \left\{ \begin{array}{ll}  |\lambda_3|^2/|\lambda_1| & {\rm if}~~~~ \lambda_1,\lambda_2,\lambda_3<0, \\ 0 & {\rm otherwise},
\end{array} \right. 
\end{eqnarray*}
where $\lambda_1, \lambda_2, \lambda_3$ are the  eigenvalues of the Hessian matrix of each voxel,
sorted so that $|\lambda_1| \geq |\lambda_2| \geq |\lambda_3|$.\\
To enhance the sensitivity of this filter to nodules of different sizes, a multi-scale approach is followed. According to the indications given in~\cite{Li,Koenderink,Lindeberg}, the $z_{dot}$ function is combined to a Gaussian smoothing at several scales $\sigma_{min}=\sigma_{1},\ldots,\sigma_{max}=\sigma_{N}$.
Within the range $\left[\sigma_{1},\sigma_{N}\right]$, intermediate smoothing scales are computed as $\sigma_{i}=r^{i-1}\sigma_{1}$ for $i=2,\ldots,N-1$, where $r=(\sigma_{N}/\sigma_{1})^{1/N-1}$. The final filter value $z_{max}$ assigned to each voxel is defined as the maximum $z_{dot}$ value obtained from the different scales (denoted $z_{dot}(\sigma_{i})$ for $i=1,\ldots,N$), multiplied by the relative scale factor:
\begin{eqnarray*}
z_{max}=\max_{i\in \{1,\ldots,N\}} \sigma_i^2 z_{dot}(\sigma_i).
\end{eqnarray*}
Once the 3D filtered matrix is calculated, a peak-detection algorithm is applied to detect the local maxima, which, sorted in decreasing order, are the list of nodule candidates identified by the filter.

\begin{figure}[ht]
\begin{center}
\includegraphics[width=.6\textwidth]{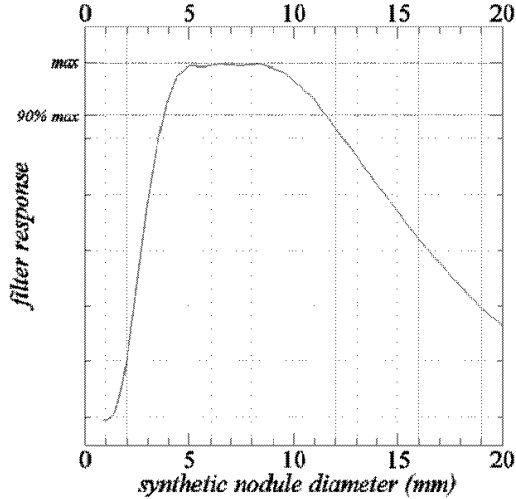}
\end{center}
\caption{Response of the multi-scale filter implemented in the scale range $\sigma_{1}$=1.25 mm, $\sigma_{N}$=2 mm and $N$=5, to synthetic nodules modeled as 3D Gaussian functions with scale $\sigma$ varying between 0.25 mm and 5~mm.}
\label{fig:range_filtro}
\end{figure}
\noindent The range $\left[\sigma_{1},\sigma_{N}\right]$ and the number $N$ of smoothing scales have to be chosen in order to make the filter able to enhance nodules of the desired dimension target.\\
Assuming a nodule can be denoted by a 3D Gaussian function of scale $\sigma$, its diameter can be reasonably assumed to be 4$\sigma$, thus accounting for more than 95\% of the nodule volume. Therefore implementing the multi-scale filter in a range $\left[\sigma_{1},\sigma_{N}\right]$ allows to enhance nodules with a diameter in a range $\left[d_{min}=\sigma_{1}*4,d_{max}=\sigma_{N}*4\right]$.\\
In figure~\ref{fig:range_filtro} the response of the filter, implemented with $\sigma_{1}$=1.25 mm, $\sigma_{N}$=2 mm and $N$=5, to synthetic nodules (3D Gaussian functions with scale $\sigma$ varying between 0.25 mm and 5 mm), is shown. As expected, the maximum filter response is obtained for nodule diameters
between $\sigma_{1}*4$=1.25$*$4 mm=5 mm and $\sigma_{N}*4$=2$*$4 mm=8 mm. However it can be noticed that the filter response remains higher than the 90\% of the maximum value for nodule diameters in all our dimension target, between 5 and 12 mm. Moreover, we found that 5 is the minimum value of $N$ required to obtain a flat response. Therefore in this work $\sigma_{1}$=1.25 mm, $\sigma_{N}$=2 mm and $N$=5 were chosen to identify nodules in our dimension target.\\
The dot-enhancement filter implemented with these parameters was run on the entire lung volume identified by the segmentation algorithm proposed in~\cite{ITK} (see figure~\ref{fig:segm} and figure~\ref{fig:segm_banda}). Then, from the filter output list, a list of \emph{internal} nodule candidates was created, constituted by filter outputs located in the previously identified lung internal region only (see figure~\ref{fig:segm_banda}). As it is explained in the next paragraph, the list contains a large number of false positives (FP).

\subsection{False positive reduction}

\subsubsection{The basic idea of the Multi-Scale Voxel-Based Neural Approach (MS-VBNA)}
Most false positive findings are crossings between blood vessels. In figure~\ref{fig:esempi_FP} some examples of FP are shown.

\begin{figure}[ht]
\begin{center}
\includegraphics[width=.9\textwidth]{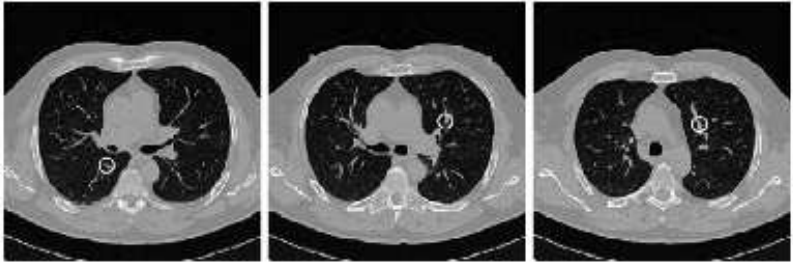}
\end{center}
\caption{Some examples of false positive findings generated by the dot-enhancement filter.}
\label{fig:esempi_FP}
\end{figure}
\noindent To reduce the number of FP/scan, we developed a procedure called \emph{Multi-Scale Voxel-Based Neural Approach} (MS-VBNA).\\
First of all, a region of interest (ROI) is defined from each internal nodule candidate of the filter output list as the set of the voxels $v$ belonging to a sphere of radius 5 pixels around the voxel identified by the filter and with intensity value $I_{v}$ above a relative threshold $t$
\begin{eqnarray*}
t=\max_{v\in sphere} I_{v}-\frac{1}{3}\biggl( \max_{v\in sphere} I_{v}-\min_{v\in sphere} I_{v}\biggr).
\end{eqnarray*}
ROIs are so defined in order to include voxels of the structures of interest (nodules or FP as, for example, blood vessel crossings) and not background voxels. In order not to have in a ROI only voxels belonging to calcific structures that could be present in the sphere, ROIs are identified only once all the voxels with $I_{v} \geq$ 200 HU are set to 200 HU. In figure~\ref{fig:pix_presi} an example of ROI corresponding to an internal nodule of diameter 5.5 mm is shown.

\begin{figure}[ht]
\begin{center}
\includegraphics[width=.9\textwidth]{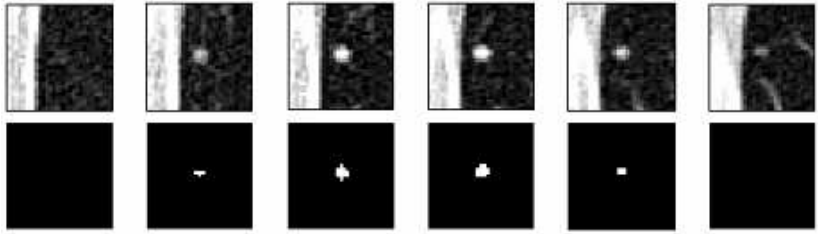}
\end{center}
\caption{Consecutive slices containing an internal nodule of diameter 5.5 mm (top) and corresponding ROI (bottom).}
\label{fig:pix_presi}
\end{figure}
\noindent At this stage we consider a nodule candidate, and consequently its ROI,  as \emph{corresponding to a nodule} (in other words, we consider a nodule \emph{found by the filter}) if the voxel identified by the filter lies within a sphere centered on the nodule and having diameter equal to the nodule dimension.\\
\\
The basic idea of the MS-VBNA is to associate to each voxel of a ROI a feature vector constituted by the intensity values of its 3D neighbors, the three eigenvalues of the \emph{gradient matrix} defined as
\begin{eqnarray*}
G_{i,j} = \left[ \sum \partial_{x_i}{I}~\partial_{x_j}{I}\right],~~~~~~~~~~~~ i,j=1,2,3,
\end{eqnarray*}
where $I(x_1,x_2,x_3)$ is the intensity function and the sums are over the neighborhood area, and the three eigenvalues of the Hessian matrix defined as
\begin{eqnarray*}
H_{i,j} = \bigl[ \partial^2_{x_i x_j}{I}\bigr],~~~~~~~~~~~~ i,j=1,2,3,
\end{eqnarray*}
where $I(x_1,x_2,x_3)$ is the intensity function~\cite{PMT} (see figure~\ref{fig:VBNA}). As we proved in~\cite{Coimbra}, using these six features, in addition to the simple voxel neighborhood rolled down into a vector, improves the system discrimination capability.

\begin{figure}[ht]
\begin{center}
\includegraphics[width=\textwidth]{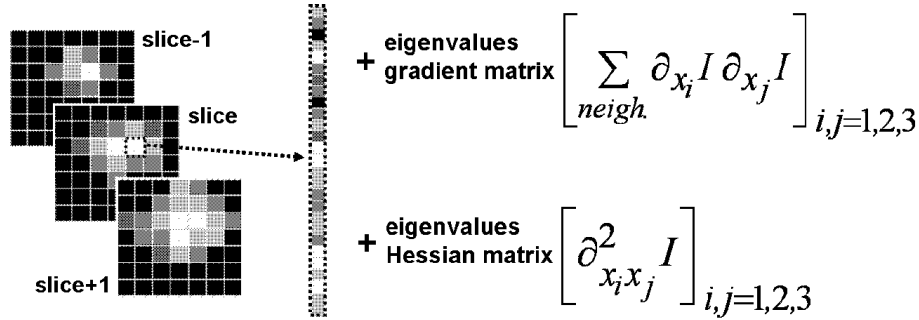}
\end{center}
\caption{Basic idea of the Voxel-Based Neural Approach to false positive reduction: each voxel is characterized by a feature vector constituted by the intensity values of its 3D neighbors and the eigenvalues of the gradient and the Hessian matrices.}
\label{fig:VBNA}
\end{figure}
\noindent Feature vectors are then classified by a standard three-layer feed-forward back-propagation neural network which is trained and tested to assign each voxel either to the nodule or normal tissue target class. 

\subsubsection{The training and testing phase}
\label{subsubsect_train_test} 
The network training and testing phase was carried out as follows.
The available dataset of 39 scans containing 75 internal nodules was partitioned into a teaching set of 15 scans containing 30 nodules and a validation set of 24 scans containing 45 nodules; the partition was defined so as to make the teaching set representative of all the nodule dimensions.\\ 
In figure~\ref{fig:isto_dim_teach_val} the distribution of nodule diameters in the teaching set and in the validation set is shown.

\begin{figure}[ht]
\begin{center}
\includegraphics[width=.7\textwidth]{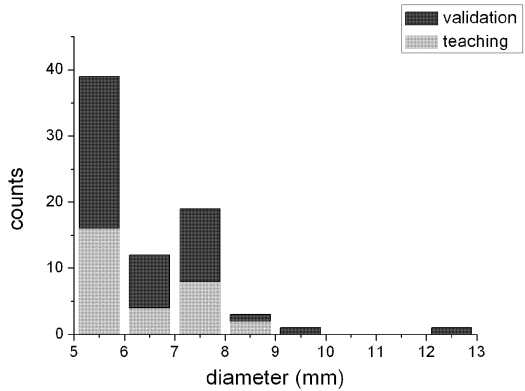}
\end{center}
\caption{Diameter distribution of the 30 internal nodules in the teaching set and the 45 internal nodules in the validation set.}
\label{fig:isto_dim_teach_val}
\end{figure}
\noindent Then, the 15 scans of the teaching set were analyzed and used to set the parameters for the training and testing phase. In particular, ROIs corresponding to the 30 nodules in the teaching set were considered. We know that clearly the number of voxels of a ROI doesn't provide a precise measurement of the nodule dimension, due to the ROI definition itself. However the mean number of voxels increases as the nodule dimension increases. In fact the mean number of voxels for nodules with diameters until 6 mm is 46, for nodules with diameters between 6 and 7 mm is 96 and for nodules with diameters above 7 mm is 124.\\
In particular, for all nodules in the teaching set the number of voxels in the ROI is greater than 20; moreover, for all but one nodules with diameters above 7 mm, the number of voxels in the ROI is greater than 100 and all but two ROIs with more than 100 voxels correspond to nodules greater than 7 mm.\\ 
Following the hypothesis that the network could really learn to recognize a nodule if the neighborhoods used to create the feature vector for the ROI voxels were large enough to intersect the nodule edge, we decided to train and test the network using two different neighborhood sizes; in particular 7$\times$7$\times$3 (7$\times$7 voxels for three consecutive slices) was chosen as \emph{small} size and 13$\times$13$\times$5 (13$\times$13 voxels for five consecutive slices) was chosen as \emph{large} size. According to our hypothesis, using the small neighborhood allowed the network to recognize at the most nodules of 7 mm of diameter, whereas for larger nodules the larger neighborhood was surely necessary.\\
Therefore, using the relation described above between the number of voxels in the ROIs and nodule dimensions, we decided to train and test the neural network using the small neighborhood size for ROIs with less than 100 voxels, the large neighborhood size for ROIs with more than 100 voxels.\\
The goal of this threshold of 100 voxels is not to define a real partition between ROIs that could be recognized using small and large neighborhoods; it is only an indicative threshold to be used to train and test the  network in an effective way.\\
It can easily be calculated that a feature vector deriving from a 7$\times$7$\times$3 neighborhood has 153 entries, 147 deriving from the neighborhood rolled down and 6 deriving from the gradient and the Hessian matrices. A feature vector deriving from a 13$\times$13$\times$5 neighborhood would have 851 features. So we decided not to consider all the neighborhood but only a down-sampled part, as shown in figure~\ref{fig:due_intorni}. In this way a feature vector deriving from a 13$\times$13$\times$5 neighborhood has also 153 entries. As a consequence, we trained and tested a feed-forward back-propagation neural network with 153 input nodes and two output nodes.

\begin{figure}[ht]
\begin{center}
\includegraphics[width=\textwidth]{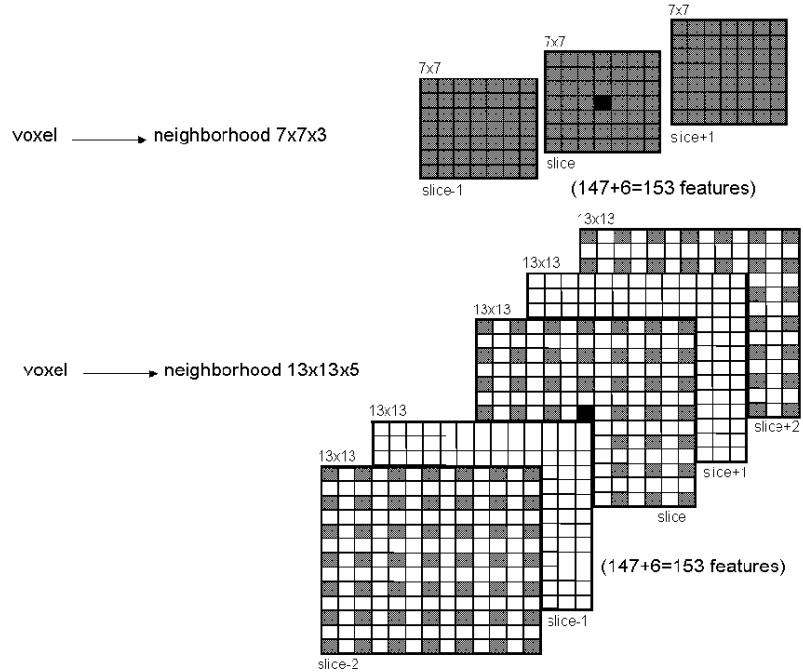}
\end{center}
\caption{Feature vectors deriving from two different neighborhood sizes: in the first case (top), the 7$\times$7$\times$3 neighborhood is rolled down into a vector of 147 entries; in the second case (bottom), the 13$\times$13$\times$5 neighborhood is down-sampled and rolled down to obtain a vector of 147 features, too. In both cases, the 6 features deriving from the gradient and the Hessian matrices are added.}
\label{fig:due_intorni}
\end{figure}
\noindent The training and testing phase was performed according to the 5$\times$2 cross validation method~\cite{Dietterich}. This method consists in performing 5 replications of the 2-fold cross validation method. In each replication, the teaching set is randomly partitioned into two sets ($A_{i}$ and $B_{i}$ for $i=1,\ldots,5$) with an almost equal number of entries. The learning algorithm is trained on each set and tested on the other one. The results achieved in each trial for the correct classification of individual voxels are reported in table~\ref{tab:5X2_n1}, where the sensitivity and the specificity values obtained on the test sets and on the whole teaching set in the ten trials are shown. Since the performance of a classifier and the comparison among different classifiers are conveniently evaluated in terms of the area $A_{z}$ under the ROC curve~\cite{Metz}, we reported in table~\ref{tab:5X2_n1} also the estimated areas under the ROC curves obtained in each trial. The average and the standard deviation of $A_{z}$ obtained in the 10 trials on testing and teaching sets are reported too, thus showing the effectiveness and the robustness of the neural classifier performance.

\begin{table}[ht]
\begin{center}
 \begin{tabular}{|c|c|c|c|c|c|c|c|}  \hline 
\multicolumn{2}{|c|}{Teaching set} &  \multicolumn{3}{|c|}{On testing set} & \multicolumn{3}{|c|}{On teaching set}\\
\hline 
Train & Test & Sens. (\%) & Spec. (\%) & $A_z$ & Sens. (\%) & Spec. (\%) & $A_z$\\
\hline 
$A_1$ & $B_1$ & 74.1 & 79.7 & 0.850 & 85.8 & 83.0 & 0.921 \\ 
$B_1$ & $A_1$ & 79.0 & 82.6 & 0.890 & 87.3 & 85.5 & 0.936 \\ 
$A_2$ & $B_2$ & 75.4 & 85.9 & 0.884 & 83.3 & 88.8 & 0.927 \\ 
$B_2$ & $A_2$ & 78.5 & 80.0 & 0.858 & 86.9 & 82.4 & 0.916 \\ 
$A_3$ & $B_3$ & 73.2 & 85.9 & 0.873 & 82.3 & 87.4 & 0.919 \\ 
$B_3$ & $A_3$ & 79.4 & 79.9 & 0.872 & 85.6 & 82.7 & 0.916 \\ 
$A_4$ & $B_4$ & 73.7 & 84.4 & 0.868 & 82.5 & 86.8 & 0.915 \\ 
$B_4$ & $A_4$ & 74.0 & 82.3 & 0.857 & 85.6 & 85.9 & 0.924 \\ 
$A_5$ & $B_5$ & 77.6 & 79.3 & 0.865 & 84.8 & 81.0 & 0.907 \\ 
$B_5$ & $A_5$ & 74.3 & 87.1 & 0.888 & 83.5 & 89.3 & 0.932 \\ 
 \hline                           
\multicolumn{2}{|c|}{Average} &  & & 0.871 &  &  & 0.921 \\ 
\multicolumn{2}{|c|}{Std deviation}&  &  & 0.014 &   &   & 0.008 \\ 	
\hline 
\end{tabular}
\end{center}
\caption{Evaluation of the performance of the standard back-propagation learning algorithm for the neural classifier according to the 5$\times$2 cross validation method.
\label{tab:5X2_n1} }
\end{table}
\noindent Among the networks with a similar performance on test sets, the second one in table~\ref{tab:5X2_n1} was more balanced with respect to sensitivity and specificity on the test set and achieved the best performance on the teaching set. Moreover it was the network with the largest area under the ROC curve, so it was expected to be the one with the greatest discrimination capability.\\
Therefore this trained neural network was applied to the ROIs voxels.

\subsubsection{The application of the trained neural network: from voxels to ROIs}
To be sure to make the approach sensible to all our nodule dimension target, we decided to apply the trained network at two different scales, characterized by two different thresholds on the number of voxels in the ROI and feature vectors deriving from the two different size voxel neighborhoods.\\ 
In particular,  for the \emph{first scale} only ROIs with more than 20 voxels are processed and for each voxel the network is applied to the feature vector deriving from the 7$\times$7$\times$3 neighborhood.\\ 
For the \emph{second scale}, only ROIs with more than 50 voxels are processed and for each voxel the network is applied to both the feature vectors that can be associated to the voxel, that is to say the feature vector deriving from the 7$\times$7$\times$3 neighborhood, obtaining the output $out_{1}=(out_{1,1},out_{1,2})$, and the feature vector deriving from the 13$\times$13$\times$5 neighborhood, obtaining the output $out_{2}=(out_{2,1},out_{2,2})$. Then the final output assigned to the voxel for the second scale is $out_{1}$ if $|out_{1,1} - out_{1,2}|>|out_{2,1} - out_{2,2}|$, $out_{2}$ otherwise.\\
This choice of the two thresholds of 20 and 50 voxels is done only depending on the analysis of the teaching set, as well as the parameters used in the training and testing phase: as explained in the subsection~\ref{subsubsect_train_test}, no nodule in the teaching set has less than 20 voxels in its corresponding ROI. Moreover, nodules that could not be recognized using only the 7$\times$7$\times$3 neighborhood have more than 50 voxels. In both cases, the choice of the threshold is conservative and it is expected not to compromise the system capability to generalize.\\ 
At each scale a nodule candidate is then classified as ``CAD nodule'' if the percentage of voxels in its ROI tagged as ``nodule'' by the neural classifier is above a threshold. By varying these thresholds at the two scales a free response receiver operating characteristic (FROC) curve can be evaluated. 

\section{RESULTS}

In figure~\ref{fig:isto_posizioni_teaching} the distribution of the positions of the 30 internal nodules of the teaching set in the lists of internal nodule candidates provided by the filter is shown.

\begin{figure}[ht]
\begin{center}
\includegraphics[width=.7\textwidth]{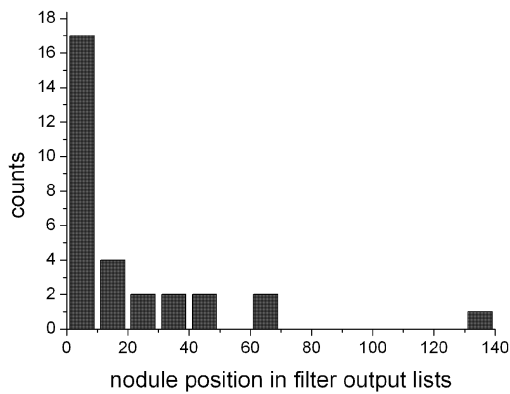}
\end{center}
\caption{Distribution of the positions of the 30 internal nodules of the teaching set in the lists provided by the filter.}
\label{fig:isto_posizioni_teaching}
\end{figure}
\noindent As it can be noticed, the lists provided by the dot-enhancement filter for the 15 scans of teaching set have to be truncated at 140 to include all annotated nodules. We fixed this parameter and we evaluated the filter sensitivity on the validation set of 24 scans containing 45 nodules, by truncating the lists at different values around it, in particular we truncated the lists at a value $M$, for $M=~$80, 100, 120, 140, 160, 180, 200.  The results are shown in table~\ref{tab:filter_sens}.

\begin{table}[t]
\begin{center}
 \begin{tabular}{|c|c|}  \hline 
Lists truncated at & Filter sensitivity (\%) \\
\hline 
80 & 88.9\\
100 & 91.1\\
120 & 91.1\\
140 & 93.3\\
160 & 95.6\\
180 & 95.6\\
200 & 95.6\\
\hline                           
\end{tabular}
\end{center}
\caption{Dot-enhancement filter sensitivity.
}
\label{tab:filter_sens} 
\end{table}
\noindent We then applied the MS-VBNA to the same truncated lists and we evaluated the performance of our CAD system in terms of FROC curves. In figure~\ref{fig:FROC}, the FROC curves obtained on the validation set by truncating the lists at $M=~$80, 140, 200 are shown. It can be noticed that the maximum sensitivity achieved by the CAD system is clearly different for different values of $M$, due to the different filter sensitivity. However, FROC curves are very close to each other up to a very good sensitivity (85\% range), thus proving the robustness of the system. In particular, a sensitivity of 86.7\% at 5.4--7.6 FP/scan and a sensitivity of 84.4\% at 4.1--5.8 FP/scan are measured. In other words, the MS-VBNA maintains a very good sensitivity (86.7\%) by eliminating 91-96\% of dot-enhancement filter false positive findings. For the slightly lower sensitivity of 84.4\%, the false positive reduction rate of the MS-VBNA is 94-97\%.

\begin{figure}[ht]
\begin{center}
\includegraphics[width=.8\textwidth]{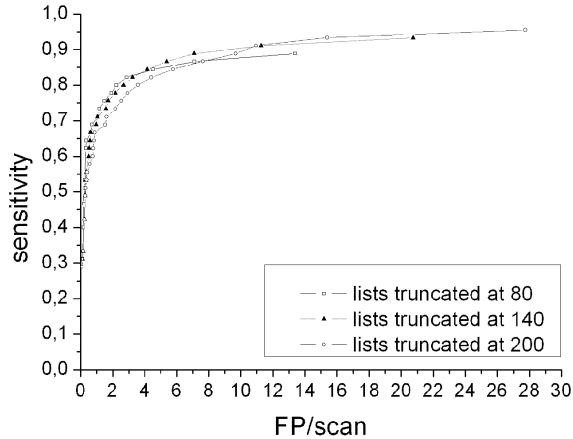}
\end{center}
\caption{FROC curves on the validation set of 24 scans containing 45 internal nodules.}
\label{fig:FROC}
\end{figure}

\section{CONCLUSIONS}
A CAD system for the identification of internal nodules with diameters greater than 5 mm was developed. The three basic modules of the system are described and the results obtained on a validation dataset of 24 low-dose CT scans with 1-mm reconstructed slice thickness containing 45 internal nodules are presented. A sensitivity of 86.7\% was obtained at a low level of false positive findings (5.4--7.6 FP/scan); the sensitivity remains high (84.4\%) even at 4.1--5.8 FP/scan. In particular, the procedure we developed for false positive reduction (MS-VBNA) works in a very satisfactory way: it eliminates more than 90\% of dot-enhancement filter false positive findings and maintains a very good sensitivity.\\
The results obtained so far are promising, but further work is foreseen.\\
The basic concept of CAD is to provide a second opinion to assist radiologists' image interpretation~\cite{CAD-DOI}. Many studies~\cite{Brochu,Yuan,Rubin,Gebhard,Wormanns} investigated and proved the CAD ability to improve the performance of radiologists in lung nodule detection in screening and clinical CT examininations.\\
For example, in~\cite{Brochu} Brochu et al. compared the performance achieved by three radiologists with different levels of experience in detecting lung nodules in 30 screening CT examinations, without the assistance of CAD and using a CAD system as second reader. The CAD system used in the study was the commercial system ImageChecker\textregistered V1.0 (R2 Technology). The CAD alone achieved a sensitivity of 79\% in the detection of nodules measuring 4 mm or larger, at a rate of 3.5 FP per examination. The sensitivity of the three radiologists in the detection of the same nodules before and after using the CAD system as second reader varied from a mean of 59\% to a mean of 90\%, with a gain of 31\%.\\
In the next phase of our work we plan to evaluate the effect of our CAD system as second reader on the performance of radiologists with different levels of experience.\\
Moreover, a validation of our CAD system against a larger database is required.
To this aim, a larger database, not only of baseline but also of repeat and follow up examinations is being collected.\\
As explained in section~\ref{subsubsect_DB}, according to ITALUNG-CT screening protocol, in follow up examinations the nodules with diameters in the 3--5~mm range should also be identified; therefore we intend to adapt our procedure, by adjusting the parameters at every step, such as the band thickness in lung internal region segmentation, the multi-scale filter range in nodule candidate identification and the voxel neighborhood size in false-positive reduction, to this new dimension target, in order to develop a CAD system useful in every phase of a screening program, on the basis of radiologists' requests.

\subsection*{Acknowledgments}
We thank Dr. L. Battolla, Dr. F. Falaschi, Dr. C. Spinelli, Radiodiagnostica 2, Azienda Ospedaliera Universitaria Pisana; Prof. D. Caramella, Dr. T. Tarantino, Diagnostic and Interventional Radiology, University of Pisa; Dr M. Mattiuzzi, Bracco Imaging S.p.A..


\begin{thebibliography}{99}
\bibitem{cancer}
J.~Ferlay, P.~Autier, M.~Boniol, M.~Heanue, M.~Colombet and P.~Boyle, \emph{Estimates of the cancer incidence and mortality in Europe in 2006}, {\emph{Annals of Oncology} {\bf 18}(3), 581--592 (2007)}.
\bibitem{cancer_stat_US}
A.~Jemal et al., \emph{Cancer Statistics, 2005}, {\emph{CA Cancer J. Clin.} {\bf 55}, 10--30 (2005)}.
\bibitem{cancer_stat_EU} 
A.~Micheli, P.~Baili, M.~Quinn, E.~Mugno, R.~Capocaccia and P.~Grosclaude (The EUROCARE Working Group), \emph{Life expectancy and cancer survival in the EUROCARE-3 cancer registry areas}, {\emph{Ann. Oncol.} {\bf 14}(5), V28--V40 (2003)}. 
\bibitem{Singh}
G.K.~Singh, B.A.~Miller and B.F.~Hankey, \emph{Changing area socioeconomic patterns in U.S. cancer mortality, 1950--1988: Part II--Lung and colorectal cancers}, {\emph{J. Natl. Cancer Inst.} {\bf 94}, 916--925 (2002)}.
\bibitem{Ihde}
D.C.~Ihde and J.D.~Minna, \emph{Non-small cell lung cancer, Part I: Biology, diagnosis, and staging}, {\emph{Curr. Probl. Cancer} {\bf 15}(2), 61--104 (1991)}.
\bibitem{Nesbitt}
J.C.~Nesbitt, J.B.~Putnam, G.L.~Walsh, J.A.~Roth and C.F.~Mountain, \emph{Survival in early-stage non-small cell lung cancer}, {\emph{Ann. Thorac. Surg.} {\bf 60}(2), 466--72 (1995)}.
\bibitem{Frost}
J.K.~Frost et al., \emph{Early lung cancer detection: results of the initial (prevalence) radiologic and cytologic screening in the John Hopkins study}, {\emph{Am. Rev. Respir. Dis.} {\bf 130}, 549--54 (1984)}.
\bibitem{Melamed}
M.R.~Melamed, B.J.~Flehinger, M.B.~Zaman, R.T.~Heelan, W.A.~Perchick and N.~Martini, \emph{Screening for early lung cancer: result of the Memorial Sloan-Kattering study in NY}, {\emph{Chest} {\bf 86}, 44--53 (1984)}.
\bibitem{Fontana}
R.S.~Fontana et al., \emph{Early lung cancer detection: results of the initial (prevalence) radiologic and cytologic screening in the Mayo Clinic study}, {\emph{Am. Rev. Respir. Dis.} {\bf 130}, 561--65 (1984)}.
\bibitem{Marcus}
P.M.~Marcus et al., \emph{Lung cancer mortality in the Mayo Lung Project: impact of extended follow-up}, {\emph{J. Natl. Cancer Inst.} {\bf{92}}, 1308--16 (2000)}.
\bibitem{Diederich}
S.~Diederich, M.G.~Lentschig, T.R.~Overbeck, D.~Wormanns and W.~Heindel, \emph{Detection of pulmonary nodules at spiral CT: comparison of maximum intensity projection sliding slabs and single-image reporting}, {\emph{Eur. Radiol.} {\bf 11},  1345 (2001)}.
\bibitem{Kaneko}
M.~Kaneko et al., \emph{Peripheral lung cancer: screening and detection with low-dose spiral CT versus radiography}, {\emph{Radiology} {\bf{201}}, 798--802 (1996)}.
\bibitem{Sone}
S.~Sone et al., \emph{Mass screening for lung cancer with mobile spiral computed tomography scanner}, {\emph{The Lancet} {\bf{351}}, 1242-45 (1998)}.
\bibitem{Itoh}
S.~Itoh et al., \emph{Lung Cancer Screening: Minimum Tube Current Required for Helical CT}, {\emph{Radiology} {\bf 215}, 175--183 (2000)}.
\bibitem{Henschke}
C.I.~Henschke et al., \emph{Early Lung Cancer Action Project: overall design and findings from baseline screening}, {\emph{The Lancet} {\bf 354}(9173), 99--105  (1999)}.
\bibitem{magic5}
R.~Bellotti et al., \emph{The MAGIC-5 project: Medical Applications on a Grid Infrastructure Connection}, {\emph{2004 IEEE Nuclear Science Symposium Conference Record} {\bf{3}}, October, 16-22, 2004, 1902--06}.
\bibitem{GRID} 
P.~Cerello et al., \emph{GPCALMA: a Grid-based Tool for Mammographic Screening}, {\emph{Methods of Information in Medicine} {\bf 44}, 244--248 (2005)}.
\bibitem{screening}
\emph{http://www.cspo.it/}
\bibitem{italung2}
G.~Picozzi et al., \emph{Screening of lung cancer with low dose spiral CT: results of a three year pilot study and design of the randomised controlled trial ``Italung-CT''}, {\emph{Radiol. Med.} {\bf 109}, 17--26 (2005)}.
\bibitem{SPIE}
I.~Gori, M.E.~Fantacci, A.~Preite Martinez and A.~Retico, \emph{An automated system for lung nodule detection in low-dose computed tomography}, {in proceedings of \emph{SPIE Medical Imaging 2007}, {\bf 6514}, 65143R (Mar. 30, 2007)}. 
\bibitem{CARSparnian}
P.~Kasae on behalf of Magic-5 Collaboration, \emph{Automated detection of pleural nodules in low-dose and thin-slice lung computed tomography}, {in proceedings of \emph{Computer Assisted Radiology and Surgery, 21th International Congress and Exhibition}, International Journal of Computer Assisted Radiology and Surgery, {\bf 2},  Supplement 1 (2007), 515}. 
\bibitem{ITK} 
J. Heuberger, A. Geissb\" uhler and H. M\" uller, \emph{Lung CT segmentation for image retrieval using the Insight Toolkit (ITK)}, {\emph{Medical Imaging and Telemedicine (MIT)}, WuYi Mountain, China, August 2005}.
\bibitem{kemerink}
G.J.~Kemerink, R.J.~Lamers, B.J.~Pellis, H.H.~Kruize and J.M.~van~Engelshoven, \emph{On segmentation of lung parenchyma in quantitative computed tomography of the lung}, {\emph{Med. Phys.} {\bf 25}(12), 2432--2439 (1998)}.
\bibitem{rolling}
S.R.~Sternberg, \emph{The Cyto-computer for morphological operations}, {\emph{Biomedical Image Processing, IEEE Computer} {\bf 16}(1),  22--34 (1983)}.
\bibitem{Li}
Q.~Li, S.~Sone and K.~Doi,  \emph{Selective enhancement filters for nodules, vessels, and airway walls in two- and three-dimensional CT scans}, {\emph{Med. Phys.} {\bf 30}(8), 2040 (2003)}.
\bibitem{Koenderink}
J.~Koenderink, \emph{The structure of image}, {\emph{Biol. Cybern.} {\bf 50}, 363--370 (1984)}.
\bibitem{Lindeberg}
T.~Lindeberg, \emph{On scale detection for different operators}, Proceedings of the \emph{Eighth Scandinavian Conference on Image Analysis}, 1993, 857--866.
\bibitem{PMT}
I.~Gori and M.~Mattiuzzi, \emph{A method for coding pixels or voxels of a digital or digitalized image : Pixel Matrix Theory (PMT)}, {\emph{European Patent Application} 05425316.6 (2005)}.
\bibitem{Coimbra}
P.~Delogu, M.E.~Fantacci, I.~Gori, A.~Preite Martinez and A.~Retico, \emph{Computer-aided detection of pulmonary nodules in low-dose CT}, {in proceedings of \emph{CompIMAGE Symposium}, October, 20--21, 2006, Coimbra, Portugal}. 
\bibitem{Dietterich} 
T.G.~Dietterich,  \emph{Approximate Statistical Test For Comparing Supervised Classification Learning Algorithms}, {\emph{Neural Computation} 1998 {\bf 10}(7), 1895--1923}. 
\bibitem{Metz} C.E.~Metz, \emph{ROC methodology in radiologic imaging}, {\emph{Invest. Radiol.} {\bf 21}(9), 720--33 (1986)}.
\bibitem{CAD-DOI} K.~Doi, \emph{Current status and future potential of computer-aided diagnosis in medical imaging }, {\emph{Br J Radiol} {\bf 78}(Special Issue), S3--S19 (2005)}.
\bibitem{Brochu} B.~Brochu et al., \emph{Computer-aided detection (CAD) of lung nodules on high-resolution MDCT screening exams}, {\emph{J. Radiol.} {\bf 88}, 573--578 (2007)}.
\bibitem{Yuan} R.~Yuan, P.M.~Vos and P.L.~Cooperberg, \emph{Computer-aided detection in screening CT for pulmonary nodules}, {\emph{AJR Am J Roentgenol} {\bf 186}, 1280--1287 (2006)}.
\bibitem{Rubin} G.D.~Rubin, et al., \emph{Pulmonary nodules on multi-detector row CT scans: performance comparison of radiologists and computer-aided detection}, {\emph{Radiology} {\bf 234}, 274--283 (2005)}.
\bibitem{Gebhard} H.H.~Gebhard, L.~Lauffer, P.~Costello and U.J.~Schoepf, \emph{Prospective evaluation of an algorithm for the automated detection of lung nodules at routine chest CT in a clinical environment}, {in proceedings of \emph{Radiological Society of North America 90th Scientific Assembly and Annual Meeting}, November 28 -- December 3, 2004, Chicago, Illinois, U.S.A.}.
\bibitem{Wormanns} D.~Wormanns, F.~Beyer, S.~Diederich, K.~Ludwig and W.~Heindel, \emph{Diagnostic performance
of a commercially available computer-aided diagnosis system for automatic detection of pulmonary nodules:
comparison with single and double reading}, {\emph{Rofo} {\bf 176}, 953--958 (2004)}.
\\


\end{thebibliography}
\end{document}